\documentclass[aps,preprintnumbers,nofootinbib,superscriptaddress]{revtex4-2}
\usepackage[utf8]{inputenc}
\usepackage{amsmath}
\usepackage{amssymb}

\def\Mpl{M_{\rm P}}

\makeatletter

\usepackage{color}
\usepackage{graphicx}
\makeatother

\begin{document}
\preprint{YITP-20-131, IPMU20-0111}
\title{Black holes in a type-II minimally modified gravity}
\author{Antonio De Felice}
\affiliation{Center for Gravitational Physics, Yukawa Institute for Theoretical
Physics,~\\
 Kyoto University, Kyoto 606-8502, Japan}
\author{Andreas Doll}
\affiliation{Department of Physics and Astronomy,~\\
 Ruprecht Karls University of Heidelberg, Heidelberg 69117, Germany}
\affiliation{Center for Gravitational Physics, Yukawa Institute for Theoretical
Physics,~\\
 Kyoto University, Kyoto 606-8502, Japan}
\author{Fran\c{c}ois Larrouturou}
\affiliation{Institut d'Astrophysique de Paris, UMR 7095, CNRS, Sorbonne Universit\'{e},~\\
 98\textsuperscript{bis} boulevard Arago, 75014 Paris, France}
\author{Shinji Mukohyama}
\affiliation{Center for Gravitational Physics, Yukawa Institute for Theoretical
Physics,~\\
 Kyoto University, Kyoto 606-8502, Japan}
\affiliation{Kavli Institute for the Physics and Mathematics of the Universe (WPI),
~\\
The University of Tokyo Institutes for Advanced Study,~\\
The University of Tokyo, Kashiwa, Chiba 277-8583, Japan}
\date{\today}
\begin{abstract}
In the context of the recently proposed type-II minimally modified gravity theory, i.e.\ a metric theory of gravity with two local physical degrees of freedom that does not possess an Einstein frame, we study spherically symmetric vacuum solutions to explore the strong gravity regime. Despite the absence of extra degrees of freedom in the gravity sector, the vacuum solutions are locally different from the Schwarzschild or Schwarzschild-(A)dS metric in general and thus the Birkhoff theorem does not hold. The general solutions are parameterized by several free functions of time and admit regular trapping and event horizons. Depending on the choice of the free functions of time, the null convergence condition may be violated in vacuum. Even in the static limit, while the solutions in this limit reduce to the Schwarzschild or Schwarzschild-(A)dS solutions, the effective cosmological constant deduced from the solutions is in general different from the cosmological value that is determined by the action. Nonetheless, once a set of suitable asymptotic conditions is imposed so that the solutions represent compact objects in the corresponding cosmological setup, the standard Schwarzschild or Schwarzschild-(A)dS metric is recovered and the effective cosmological constant agrees with the value inferred from the action. 
\end{abstract}
\maketitle

\section{Introduction}

While general relativity (GR) remains so far the best description of gravity in a wide range of scales, there are a number of reasons to explore possible modifications. One motivation is the hope to address mysteries in the universe such as the origins of dark energy, dark matter and cosmic inflation. It would be interesting if some of those puzzles in the modern cosmology could be restated in different ways by modifying gravity so that one could approach them from different physical viewpoints. Another motivation is to help constructing a theory of quantum gravity. Indeed, it is believed that GR should be modified at least at short distances, in order to resolve theoretical inconsistencies between GR and quantum theory. A yet another motivation comes from our hope to understand GR itself. Even if GR is the genuine description of gravity all the way from a distance scale somewhat above the Planck length up to cosmological scales, the only possible way to prove it experimentally or/and observationally is to predict possible deviations from GR and then to constrain such deviations by experiments or/and observations. In fact, if we are to believe the latest cosmological data, little hope is left for the concordance model of cosmology to be correct.

In the landscape of various theories of gravity beyond GR, there are (at least) three important theoretical checkpoints on each theory. What are the physical degrees of freedom in the gravity sector? How do they interact with themselves and with the matter sector? What is the regime of validity of the theory? Regarding the first question, given the recent observations of gravitational waves, we know that the minimal number of local physical degrees of freedom in the gravity sector is two, corresponding to the tensorial gravitational waves~\cite{Abbott_2019}. It is therefore important and interesting to ask whether this minimal number can be saturated by theories beyond GR. Alternative gravity theories with two local physical degrees of freedom in the gravity sector were recently dubbed minimally modified gravity (MMG) theories~\cite{Lin:2017oow} and then developed~\cite{Aoki:2018zcv,Aoki:2018brq,Mukohyama:2019unx,Carballo-Rubio:2018czn,DeFelice:2020eju,Aoki:2020lig,Aoki:2020oqc,DeFelice:2020cpt} (see \cite{DeFelice:2015hla,DeFelice:2015moy,Bolis:2018vzs,DeFelice:2018vza,Afshordi:2006ad,Iyonaga:2018vnu,Feng:2019dwu,Gao:2019twq} for other examples of MMG theories).

As argued in \cite{Aoki:2018brq}, all MMG theories can be classified into type-I and type-II. A type-I MMG theory has an Einstein frame, in which all GR solutions are solutions at least locally if the matter sector is minimally coupled to the metric, and thus gravity in type-I MMG deviates from GR only due to non-trivial matter coupling. Hence, one can systematically generate all type-I MMG theories from GR, following the prescription of \cite{Aoki:2018brq}, i.e.\ by a canonical transformation followed by the addition of a gauge-fixing condition and coupling to the matter sector. On the other hand, a type-II MMG theory does not have an Einstein frame. While an example of type-II MMG in the context of massive gravity has been known~\cite{DeFelice:2015hla,DeFelice:2015moy,Bolis:2018vzs,DeFelice:2018vza}, a systematic construction method of all type-II MMG theories has not yet been developed.

Recently, a new example of type-II MMG was found in \cite{DeFelice:2020eju}. One starts with the Hamiltonian of GR, performs a canonical transformation to a new frame, adds a gauge-fixing term and a cosmological constant in the new frame, and then goes back to the original frame by the inverse canonical transformation. In this way, one obtains the Hamiltonian of the theory. After that, it is straightforward to perform a Legendre transformation to obtain the action of the gravity sector. One can then add matter fields that are minimally coupled to the metric in the original frame.
The price to pay to have a non-GR theory that propagates only 2 gravitational degrees of freedom is to break the temporal diffeomorphism, which is made explicit by the gauge fixing term.
The gravity sector of the theory is characterized by a potential function $V(\phi)$ of an auxiliary three-dimensional scalar field $\phi$. Despite the absence of extra local physical degrees of freedom, one can reproduce any expansion history of the homogeneous and isotropic background universe for a given matter sector by arranging the form of the potential $V(\phi)$ properly, as far as the matter sector strictly satisfies the null energy condition at the level of the background. (On the other hand, in the absence of matter but with or without a cosmological constant in the original frame, the background universe is either de Sitter or Minkowski.) Once the form of $V(\phi)$ is fixed, one can also derive the evolution equation of cosmological perturbations without any ambiguities. Conversely, from the observed behavior of the cosmological background and perturbations around it, the form of the potential can be constrained. Therefore, this setup can be tested against cosmological data and might help reducing some tensions indicated by cosmological probes~\cite{DeFelice:2020cpt}.

The purpose of the present paper is to explore the strong gravity regime of this new type-II MMG theory by studying black hole solutions. For simplicity we restrict our consideration to spherically symmetric configurations but allow for time-dependence in the metric and the auxiliary field. 

The rest of the present paper is organized as follows. Sec.~\ref{sec:model_description} introduces the theory and the ansatz considered in this work. Sec.~\ref{sec:static} is then devoted to the construction of its static spherically symmetric solutions. The time-dependent solutions are then derived and discussed in Sec.~\ref{sec:time_dep}, before concluding our work in Sec.~\ref{sec:summary}.

\section{Model description}\label{sec:model_description}

\subsection{A type-II MMG in a nutshell}

The gravitational model analyzed in this paper is a theory of type-II minimally modified gravity (MMG) that was introduced in \cite{DeFelice:2020eju}. It is a modification of standard GR that adds an auxiliary three-dimensional scalar field but that does not change the number of physical degrees of freedom. Throughout \cite{DeFelice:2020eju} and the present paper, we adopt the Arnowitt-Deser-Misner (ADM) decomposition of spacetime: 
\begin{equation}
\text{d}s^2 =-N^{2}\text{d}t^{2}+\gamma_{ij}(\text{d}x^{i}+N^{i}\text{d}t)(\text{d}x^{j}+N^{j}\text{d}t)\,,\label{eqn:metric-ADM}\\ 
\end{equation}
where $N$, $N^i$ and $\gamma_{ij}$ are the lapse function, the shift vector and the spatial metric, respectively. As outlined in the introduction, the construction of the type-II MMG then follows a simple recipe starting with the Hamiltonian of GR without any matter,
\begin{equation}
H_{{\rm tot}} =\int d^{3}x[N\mathcal{H}_{0}(\gamma,\pi)+N^{i}\mathcal{H}_{i}(\gamma,\pi)+\lambda\pi_{N}+\lambda^{i}\pi_{i}]\,, 
\end{equation}
where $\{\pi_N,\pi_i,\pi^{ij}\}$ are conjugate to $\{N,N^i,\gamma_{ij}\}$, and
\begin{align}
\mathcal{H}_{0} & =\frac{2}{\Mpl^{2}\sqrt{\gamma}}\left(\gamma_{ik}\gamma_{jl}-\frac{1}{2}\,\gamma_{ij}\gamma_{kl}\right)\pi^{ij}\pi^{kl}-\frac{\Mpl^{2}\sqrt{\gamma}}{2}\,R(\gamma)\,,\label{eqn:H0-GR}\\
\mathcal{H}_{i} & =-2\sqrt{\gamma}\gamma_{ij}D_{k}\!\left(\frac{\pi^{jk}}{\sqrt{\gamma}}\right),\label{eqn:Hi-GR}
\end{align}
are the Hamiltonian and momentum constraints, respectively, $\lambda$ and $\lambda^{i}$ are Lagrange multipliers, and $D_{k}$ is the spatial covariant derivative compatible with the spatial metric $\gamma_{ij}$. The Hamiltonian is then canonically transformed into a different frame via a generating functional that depends on the old momenta and new variables $\{\mathfrak{N},\mathfrak{N}^i,\Gamma_{ij} \}$: 
\begin{equation}
\mathcal{F}=-\int d^{3}x\,[\Mpl^{2}\sqrt{\Gamma}f(\phi,\psi)+\mathfrak{N}^{i}\,\pi_{i}]\,,
\end{equation}
where
\begin{equation}
 \phi=\frac{1}{\Mpl^{2}\sqrt{\Gamma}}\,\pi^{ij}\,\Gamma_{ij}\,,\qquad\psi=\frac{1}{\Mpl^{2}\sqrt{\Gamma}}\,\pi_{N}\,\mathfrak{N}\,.
\end{equation}
This form of the generating functional was chosen in \cite{DeFelice:2020eju} as to make the calculations simpler. We continue to use this functional in order to keep the same simplicity. Though the functions $\phi$ and $\psi$ can be expressed in terms of the variables and momenta, in the following they are kept as auxiliary independent scalar fields in three spatial dimensions. They can depend on time but they are not four dimensional scalars.

In this new frame, a cosmological constant $\tilde{\Lambda}$ is added as well as a gauge fixing term that keeps the degrees of freedom of the model at only two. Afterwards, the inverse canonical transformation is applied to this modified Hamiltonian to put it back into its original frame. We now arrive at a model that is no longer GR but still has only two degrees of freedom:
\begin{align}
H_{{\rm tot}}=\int d^{3}x & \left[N\mathcal{H}_{0}(\gamma,\pi)+N^{i}\mathcal{H}_{i}(\gamma,\pi)+\lambda\,\pi_{N}+\lambda^{i}\pi_{i}+\sqrt{\gamma}\lambda_{C}\left(\phi-\frac{f_{\phi}^{1/2}}{\Mpl^{2}}\,\frac{\pi^{ij}}{\sqrt{\gamma}}\,\gamma_{ij}\right)\right.\nonumber \\
 & {}+\left.\sqrt{\gamma}\lambda_{D}\left(\psi-\frac{f_{\phi}^{3/2}}{\Mpl^{2}f_{\psi}}\,\frac{\pi_{N}}{\sqrt{\gamma}}\,N\right)+\lambda_{\phi}\pi_{\phi}+\lambda_{\psi}\pi_{\psi}
 +\sqrt{\gamma}\,\lambda_{{\rm gf}}^{i}\,\partial_{i}\phi+\frac{\Mpl^{2}}{f_{\psi}f_{\phi}^{3/2}}\,N\sqrt{\gamma}\tilde{\Lambda}\right]\,,\label{eqn:Htot_pre}
\end{align}
where we have naturally denoted $f_\phi \equiv \frac{\partial f}{\partial\phi}$ and $f_\psi \equiv \frac{\partial f}{\partial\psi}$. 
For a full proof of the two degrees of freedom being the only ones present please refer to the paper \cite{DeFelice:2020eju}.

An analysis of the dynamical constraints of the system reveals that terms second or higher order in $\psi$ in the function $f(\phi,\psi)$ do not play any roles. It thus suffices to expand $f(\phi,\psi)$ up to first order in $\psi$ as
\begin{equation}
f(\phi,\psi)=\bar{f}(\phi)+f_{1}(\phi)\psi\,,
\end{equation}
This leads to the final Hamiltonian: 
\begin{align}
H_{{\rm tot}} & =\int d^{3}x\!\left[N\mathcal{H}_{0}(\gamma,\pi)+N^{i}\mathcal{H}_{i}(\gamma,\pi)+\lambda\,\pi_{N}+\lambda^{i}\pi_{i}+\sqrt{\gamma}\lambda_{C}\!\left(\phi-\frac{f_{0}^{1/2}}{\Mpl^{2}}\,\frac{\pi^{ij}}{\sqrt{\gamma}}\,\gamma_{ij}\right)\right.\nonumber \\
 & \left.{}+\lambda_{\phi}\pi_{\phi}+\sqrt{\gamma}\lambda_{{\rm gf}}^{i}\,\partial_{i}\phi+\frac{1}{f_{1}f_{0}^{3/2}}\,N\sqrt{\gamma}\Mpl^{2}\tilde{\Lambda}\right],
\end{align}
with $f_0 \equiv \frac{\text{d} \bar{f}}{\text{d}\phi}$.

The Lagrangian for this theory is obtained by performing a Legendre transformation.
Relabeling the Lagrange multipliers, it is written as: 
\begin{equation}
\mathcal{L}=\frac{\Mpl^{2}}{2}\,N\sqrt{\gamma}\Bigg[R+K_{ij}\,K^{ij}-K^{2}-2V(\phi)-2\,\frac{\lambda_{{\rm gf}}^{i}}{N}\,\partial_{i}\phi-\frac{3\lambda^{2}}{2}-2\lambda\,(K+\phi)\Bigg]\,,\label{eq:Lagr}
\end{equation}
where we have defined $V(\phi) = \tilde{\Lambda}/f_{1}f_{0}^{3/2}$ and the extrinsic curvature is 
\begin{equation}
K_{ij}=\frac{1}{2N}(\dot{\gamma}_{ij}-D_{i}N_{j}-D_{j}N_{i})\,,
\end{equation}
and $K = \gamma^{ij}K_{ij}$ is its trace.
As the final step, the standard matter terms can be added with a minimal coupling to the metric \eqref{eqn:metric-ADM}.

The theory described by the action $I=\int dtd^3\vec{x}\mathcal{L}$, where $\mathcal{L}$ is given by (\ref{eq:Lagr}), breaks a part of the full four-dimensional diffeomorphism invariance and is invariant under only the so-called foliation-preserving diffeomorphism, which is a combination of the space-independent time reparametrization and the spatial diffeomorphism, 
\begin{equation}
 t \to t'(t)\,, \quad x^i \to {x^i}'(t,x^j)\,.  \label{eqn:foliation-preserving-diffeo}
\end{equation}

Apart from constructing the model and counting its degrees of freedom in \cite{DeFelice:2020eju}, the equations of motion of a flat Friedmann–Lema\^{i}tre–Robertson–Walker (FLRW) background metric were also calculated. The resulting conservation equation, Friedmann equation and acceleration equation read: 
\begin{gather}
\frac{\dot{\rho}}{N}+3H(\rho+P)=0\,,\qquad3\Mpl^{2}H^{2}=\rho+\rho_{\phi}\,,\label{eqn:Friedman}\\
2\Mpl^{2}\frac{\dot{H}}{N}=-(\rho+P)-(\rho_{\phi}+P_{\phi})\,,
\end{gather}
where 
\begin{align}
\rho_{\phi} & \equiv \Mpl^{2}(V-\phi V_{,\phi})+\frac{3}{4}\Mpl^{2}V_{,\phi}^{2}\,,\\
P_{\phi} & \equiv-\frac{3}{2}(\rho+P)V_{,\phi\phi}-\rho_{\phi}\,.
\end{align}
On the other hand, the Lagrange multipliers $\lambda$ and $\lambda_{{\rm gf}}^{i}$ as well as the auxiliary field $\phi$ for the FLRW background are 
\begin{equation}
 \lambda=-\frac23\,\phi-2H\,, \quad \lambda_{{\rm gf}}^{i} =0\,, \quad \phi=\frac32\,V_{,\phi}-3H\,. \label{eqn:lambda-lambdagf-phi-FLRW}
\end{equation}

Furthermore, it is possible to reconstruct the potential $V(\phi)$ when only the FLRW background dynamics is given. The only assumption one has to make is that the total stress energy tensor satisfies the condition $\rho+P>0$, and that the universe should be expanding, i.e.\ $H>0$. This ensures that $\phi$ is a steadily increasing function according to the equation of motion for $\phi$:
\begin{equation}
\frac{d\phi}{d\mathcal{N}}=\frac{3}{2}\frac{\rho+P}{\Mpl^2 H}\,.\label{eqn:dphidN}
\end{equation}
Here, $\phi(\mathcal{N})$ is a function of the e-fold variable $\mathcal{N}=\ln(a/a_{0})$. Because of its steady increase, $\phi(\mathcal{N})$ is invertible so that $\mathcal{N}=\mathcal{N}(\phi)$ exists and one can calculate the potential $V$ from the first Einstein equation: 
\begin{equation}
V=\frac{1}{3}\phi^{2}-\frac{\rho\bigl(\mathcal{N}(\phi)\bigr)}{\Mpl^{2}}\,.
\end{equation}

\subsection{Spherically symmetric ansatz and basic equations}

In the following, we will study the strong-field regime of the theory~\eqref{eq:Lagr}, in spherically symmetric configurations. 
This allows us to use the ansatz
\begin{equation}
N=N(t,r)\,,\quad N_{i}\text{d}x^{i}=B(t,r)F(t,r)\text{d}r\,,\quad\gamma_{ij}\text{d}x^{i}\text{d}x^{j}=F(t,r)^{2}\text{d}r^{2}+r^{2}\text{d}\Omega^2\,,
\end{equation}
where $\text{d}\Omega^2$ is the usual metric of the unit $2$-sphere, so that $B$ is the tetrad component of the shift vector and that the $4$-dimensional metric is 
\begin{equation}
\text{d}s^{2}=-N(t,r)^{2}\text{d}t^{2}+\left[F(t,r)\text{d}r+B(t,r)\text{d}t\right]^{2}+r^{2}\text{d}\Omega^{2}\,.\label{eq:ansatz}
\end{equation}
As for the auxiliary fields, spherical symmetry fixes the ansatz
\begin{equation}
 \phi = \phi(t,r)\,,\qquad\lambda = \lambda(t,r)\,,\qquad
  \lambda_{{\rm gf}}^{i}\,\partial_{i}=\lambda^{r}(t,r)\,\frac{\partial_{r}}{F(t,r)}\,.
\label{eq:ansatz_aux}
\end{equation}

\section{Static solution}\label{sec:static}

Let's first focus on static configurations in vacuum, i.e.\ implementing the ansatz~\eqref{eq:ansatz}-\eqref{eq:ansatz_aux} with time-independent quantities.
With this ansatz, only one gauge-fixing constraint remains, namely
\begin{equation}
\mathcal{E}_{\lambda^{r}}=-M_{\text{Pl}}^{2}r^{2}F\,\partial_{r}\phi\,,
\end{equation}
thus $\phi(r)=\phi_{0}$ is constant.
The non-dynamical fields $\lambda$ and $\phi$ obey the equations of motion 
\begin{equation}
\mathcal{E}_{\lambda}=-\frac{3M_{\text{Pl}}^{2}r^{2}NF}{2}\left[\lambda+\frac{2\phi_{0}}{3}-\frac{2}{3r^{2}NF}\,\partial_{r}\left(r^{2}B\right)\right]\quad\text{and}\quad\mathcal{E}_{\phi}=M_{\text{Pl}}^{2}\left[\partial_{r}\left(r^{2}\lambda^{r}-\frac{2r^{2}B}{3}\right)+r^{2}NF\left(\frac{2\phi_{0}}{3}-V'_{0}\right)\right]\,,\label{eq:static_eom_lambdaphi}
\end{equation}
where we naturally shortened $V_{0}'=V'(\phi_{0})$. Those equations allow us to write $\lambda$ and $\lambda^{r}$ in terms of $N$, $B$ and $F$. The equation of motion for the shift $B$ is of the form 
\begin{equation}
\mathcal{E}_{B}=-\frac{2M_{\text{Pl}}^{2}}{3rF}\,\partial_{r}\left[\frac{r^{4}}{NF}\partial_{r}\left(\frac{B}{r}\right)\right]\,,\label{eq:static_eom_B}
\end{equation}
so we can express $B=3\kappa_{0}r\int_{r_{0}}^{r}\text{d}uNFu^{-4}$, where $r_{0}$ and $\kappa_{0}$ are constants of integration. Finally the equations of motion for $N$ and $F$ are solved by 
\begin{equation}
F(r)=\frac{N_{0}}{N(r)}=\left(1-\frac{2\mu_{0}}{r}-\frac{\Lambda_{0}r^{2}}{3}+\frac{\kappa_{0}^{2}}{r^{4}}\right)^{-1/2}\,,\label{eq:N_sol}
\end{equation}
where $\mu_{0}$ is a constant of integration and we have defined $\Lambda_{0}=V_{0}-\phi_{0}^{2}/3$. The constant $N_{0}$ can be set to any positive value by the space-independent time reparametrization, which is a part of the foliation-preserving diffeomorphism (\ref{eqn:foliation-preserving-diffeo}), and it is convenient to set it to unity as we shall see later on.

So finally, the static and spherically symmetric solutions of our theory are parameterized by 5 constants $\{\mu_{0},b_{0},\kappa_{0},\phi_{0},\ell_{0}\}$ as 
\begin{subequations}
\label{eq:static_sol} 
\begin{align}
 & \text{d}s^{2}=-\frac{\text{d}t^{2}}{F^{2}(r)}+\left[F(r)\,\text{d}r+\left(b_{0}r-\frac{\kappa_{0}}{r^{2}}\right)\text{d}t\right]^{2}+r^{2}\text{d}^{2}\Omega\,,\label{eq:static_sol_ds2}\\
 & \phi=\phi_{0}\,,\qquad\lambda=2b_{0}-\frac{2\phi_{0}}{3}\,,\qquad\lambda_{{\rm gf}}^{i}\,\partial_{i}=\left[\frac{\ell_{0}}{r^{2}}+\left(V'_{0}+2b_{0}-\frac{2\phi_{0}}{3}\right)\frac{r}{3}\right]\frac{\partial_{r}}{F(r)}\,.\label{eq:static_sol_aux}
\end{align}
\end{subequations}
where $F$ is defined in (\ref{eq:N_sol}).
We have also defined $b_{0}=\kappa_{0}/r_{0}^{3}$ and $\ell_{0}$ is the constant of integration associated with the equation of motion for $\phi$~\eqref{eq:static_eom_lambdaphi}. When setting $b_{0}$ and $\kappa_{0}$ to zero, this family of metrices reduces to the usual Schwarzschild-de Sitter metrices with mass $\mu_{0}$ and cosmological constant $\Lambda_{0}$.

\subsection{Horizon regularity}

Scalars made of the intrinsic and extrinsic curvatures for the family of metrices~\eqref{eq:static_sol_ds2} are given by 
\begin{subequations}
\begin{align}
 & R_{ijkl}R^{ijkl}=\frac{4\Lambda_{0}^{2}}{3}+\frac{24\mu_{0}^{2}}{r^{6}}+\frac{8\kappa_{0}^{2}}{r^{6}}\left(\Lambda_{0}-\frac{6\mu_{0}}{r^{3}}+\frac{9\kappa_{0}^{2}}{2r^{6}}\right)\,,\\
 & 
 R_{ij}R^{ij}=\frac{4\Lambda_{0}^{2}}{3}+\frac{6\mu_{0}^{2}}{r^{6}}+\frac{8\kappa_{0}^{2}}{r^{6}}\left(\Lambda_{0}-\frac{3\mu_{0}}{2r^{3}}+\frac{9\kappa_{0}^{2}}{4r^{6}}\right)\,,\qquad
  R=2\Lambda_{0}+\frac{6\kappa_{0}^{2}}{r^{6}}\,,\label{eq:static_R3D}\\
 & 
 K_{ij}K^{ij}=3b_{0}^{2}+\frac{6\kappa_{0}^{2}}{r^{6}}\qquad\text{and}\qquad K=-3b_{0}\,.
\end{align}
\label{eq:static_physical_qtt}
\end{subequations}
They are well-behaved in all space but the origin $r=0$ : there is no geometrical singularity at the horizons.

\subsection{Black hole mass and effective cosmological constant}

The family of metrices~\eqref{eq:static_sol_ds2} describes a set of Einstein space-times, as their associated 4-dimensional Ricci tensor read 
\begin{equation}
R_{\mu\nu}=\Lambda_\text{eff} \, g_{\mu\nu}\,, \quad  
 \Lambda_\text{eff} = \Lambda_0 + 3 b_0^2\,.
\end{equation}
This indicates that those space-times bear an effective cosmological constant $\Lambda_{\text{eff}}=\Lambda_{0}+3b_{0}^{2}$. Moreover, thanks to Birkhoff theorem, this implies that the $4$-dimensional metric is locally Schwarzschild-de Sitter ($\Lambda_{\text{eff}}>0$), Schwarzschild ($\Lambda_{\text{eff}}=0$), or Schwarzschild-AdS ($\Lambda_{\text{eff}}<0$). In order to determine the mass of the black hole, let us compute the (generalized) Misner-Sharp mass. For a spherically symmetric metric of the form 
\begin{equation}
 \text{d}s^2 = h_{ab} \text{d}x^a \text{d}x^b + r^2 \text{d}\Omega^2\,,
\end{equation}
where $h_{ab}\text{d}x^a\text{d}x^b$ is the metric of a $2$-dimensional spacetime spanned by the $2$-dimensional coordinates $x^c$ ($c=0,1$), often called the orbit space, and $r$ is a non-negative function of the coordinates of the orbit space, the (generalized) Misner-Sharp mass $M$ is defined by 
\begin{equation}
 h^{ab}\partial_a r\partial_b r = 1 - \frac{2M}{r} -\frac{\Lambda_{\text{eff}}}{3}r^2\,.
  \label{eqn:def-Misner-Sharp-mass}
\end{equation}
Here, $h^{ab}$ is the inverse of $h_{ab}$. For the family of metrices~\eqref{eq:static_sol_ds2}, a straightforward calculation results in 
\begin{equation}\label{eq:expr-Misner-Sharp-mass}
 M = \mu_0 - \kappa_0 b_0\,.
\end{equation}
Therefore, the mass and cosmological constant of our Schwarzschild-de Sitter black holes are different from $\mu_0$ and $\Lambda_0$, respectively.

The fact that $\kappa_{0}$ does not enter $\Lambda_{\text{eff}}$ can be understood from the fact that this parameter is linked to a coordinate change. Starting with a usual Schwarzschild-de Sitter space-time with the mass $\mu_{0}$ and the cosmological constant $\Lambda_{0}$, and performing the time redefinition $t\rightarrow t-\kappa_{0}\,\int\frac{F^{3}(r)r^{2}\,\text{d}r}{\kappa_{0}^{2}F^{2}(r)-r^{4}}$, the line element reads 
\begin{equation}
\text{d}s_{\text{Sch-dS}}^{2}=-\left(1-\frac{2\mu_{0}}{r}-\frac{\Lambda_{0}r^{2}}{3}\right)\text{d}t^{2}+\frac{\text{d}r^{2}}{1-\frac{2\mu_{0}}{r}-\frac{\Lambda_{0}r^{2}}{3}+\frac{\kappa_{0}^{2}}{r^{4}}}-\frac{2\kappa_{0}}{r^{2}}\frac{\text{d}t\,\text{d}r}{\sqrt{1-\frac{2\mu_{0}}{r}-\frac{\Lambda_{0}r^{2}}{3}+\frac{\kappa_{0}^{2}}{r^{4}}}}+r^{2}\text{d}\Omega^{2}\,,
\end{equation}
which is nothing but the line element~\eqref{eq:static_sol_ds2} with $b_{0}=0$. As $\kappa_{0}$ is only associated to a boost, it is natural that it does not enter the effective cosmological constant. Let's nevertheless recall that while such a parameter would be unphysical in GR, this is not the case in the framework of this study: due to the breaking of temporal diffeomorphisms, two solutions with different $\kappa_{0}$ are physically distinct, as the applied time redefinition does not enter the class~\eqref{eqn:foliation-preserving-diffeo}.

The $b_0$ parameter represents the leftover freedom in spacetime slicing. Indeed, as shown in~\eqref{eq:static_physical_qtt}, it does not enter $R_{ijkl}$, but only the extrinsic curvature, when $\kappa_0$ only enters $R_{ijkl}$.

\subsection{Matching to cosmology}

In the empty cosmological setup, in general a de Sitter solution in the flat FLRW form is present, as $H^2=H_0^2=\rho_\phi(\phi_0)/(3\Mpl^2)\equiv\Lambda_{\rm eff}/3=b_0^2 + \Lambda_0/3$, which sets the value of $b_0^2$~\cite{DeFelice:2020eju}. The $4$-dimensional metric of the de Sitter solution in the flat slicing is
\begin{equation}
 \text{d}s^2 = -dt^2 + e^{2H_0t}(d\rho^2 + \rho^2\text{d}\Omega^2)\,.
\end{equation}
Since the theory enjoys the spatial diffeomorphism invariance, the metric is physically equivalent to 
\begin{equation}
 \text{d}s^2 = -dt^2 + (dr-H_0rdt)^2 + r^2\text{d}\Omega^2\,,\label{eq:match_cosmo}
\end{equation}
which is spatially flat and manifestly static. Here, $r=e^{H_0t}\rho$.

If one requires that the Schwarzschild-de Sitter black holes asymptotically matches with this de Sitter solution, then the intrinsic curvature should vanish at infinity. By means of~\eqref{eq:static_R3D} it implies that $\Lambda_0 = 0$, and thus $V(\phi_0) = \phi_0^2/3$. In this case, it is convenience to set $\lim_{r\to\infty}N=1$ (setting $N_0=1$ in Eq.\ (\ref{eq:N_sol}), as $F\to1$) by the space-independent time reparametrization, which is a part of the foliation-preserving diffeomorphism (\ref{eqn:foliation-preserving-diffeo}), so that one can match the metric (\ref{eq:static_sol_ds2}) to the metric (\ref{eq:match_cosmo}) at infinity not only up to the foliation-preserving diffeomorphism (\ref{eqn:foliation-preserving-diffeo}) but also explicitly. The matching between the two metrics then requires $b_0=-H_0$ as well. In this same case, the auxiliary fields $\phi$ and $\lambda$ are also well matching their cosmological values: $\phi$ is constant, which agrees with its homogeneous behavior in cosmology and $\lambda=-2(\phi+K)/3$ (see (\ref{eqn:lambda-lambdagf-phi-FLRW})). The gauge fixing field $\lambda_{{\rm gf}}^{i}$ should vanish at spatial infinity, which is realised if $2b_0 + V_0'-2\phi_0/3=0$, which agrees with (\ref{eqn:lambda-lambdagf-phi-FLRW}).

In the above we have imposed the exact matching to the de Sitter solution with the flat FLRW slicing. In a more realistic situation, the boundary condition within the spherically symmetric ansatz may/should be modified one way or another to the extent that the would-be mismatch between the cosmological solution and the black hole solution can be absorbed by physical effects in the intermediate region (such as deviations from spherical symmetry, the existence of interstellar matter, etc.) and/or by non-trivial cosmology (e.g. the existence of cold dark matter, baryons, radiation, etc.). Nonetheless, the properties of the black hole solution at astrophysical scales are expected to be insensitive to such modifications of the boundary condition as far as there is a large enough hierarchy between the size of the black hole and the scales associated with the physical effects in the intermediate and cosmological scales.

Let's finally note that if one takes the limit $b_0\to 0$ (after matching to the flat FLRW de Sitter solution), then the Schwarzschild-de Sitter solution reduces to the asymptotically flat Schwarzschild solution.

\subsection{Summary of static vacuum BH solution}

In a nutshell, the static and spherically symmetric solutions of our theory are given by Schwarzschild-de Sitter spacetimes~\eqref{eq:static_sol_ds2}, parameterized by $4$ parameters: the mass $M$, the effective cosmological constant $\Lambda_\text{eff}$, and the two slicing parameters $b_0$ and $\kappa_0$, linked to the way the spacetime is foliated by constant time hypersurfaces. The two other parameters $\Lambda_0$ and $\mu_0$ are not independent of the $4$ parameters mentioned above, and are defined as 
\begin{equation}
 \mu_0 = M +\kappa_0 b_0
 \qquad \text{and} \qquad
\Lambda_0 = \Lambda_\text{eff} - 3 b_0^2\,.
\end{equation}

As for the auxiliary sector~\eqref{eq:static_sol_aux}, it is parameterized by 2 additional constants: the value of the auxiliary field $\phi$, and a integration constant entering the gauge-fixing field.

Those solutions are perfectly asymptotically matching the de Sitter solutions found in~\cite{DeFelice:2020eju} as long as $V(\phi_0) = \phi_0^2/3$ (\emph{ie.} $\Lambda_0 = 0$) and $b_0=-H_0=\phi_0/3-V_0'/2$.

\subsection{Exterior solution of a static star}

So far, we have dealt with solutions which represent static black holes in vacuum and we were able to fix some of the free parameters of the solution by their asymptotic behavior at infinity. Obviously, for a black hole solution one cannot impose a boundary condition at $r=0$ since it is a singularity. The situation is different for stars. 

Instead of black holes, in this subsection we thus briefly consider static solutions that represent spherically symmetric, static stars. While one can still impose the same asymptotic behavior at infinity, in this case it is important to understand the physics around the origin, i.e.\ for $r\to0$. If the vacuum solution has to correspond to the solution of the spacetime outside a compact object, then we need to model a star in this theory. For example, we would need to introduce a perfect fluid with some given equation of state. In general exact matter solutions are difficult to find, however their interior solution once matched with the exterior solution will fix the value of $\mu_0$, $\ell_0$ and $\kappa_0$ in general.

Let us consider first, for the sub-case $B=0$, the interior solution of a compact object composed of a general matter fluid. Notice that this case is consistent with one of the boundary conditions to be set in the origin in order to avoid singularities, namely $\lim_{r\to0} B(r)/r={\rm const}$. In this case, one finds the following solution for the interior metric radial function $F$,
\begin{equation}
  F^{-2}=1+\frac{C_1}{r}-\frac{\int^rdR\,R^2\,\rho(R)}{r\Mpl^2}-\frac13\,\Lambda_0\,r^2\,,
\end{equation}
and we need to set the constant $C_1=0$ for regularity. Then it should be noted that in this solution no term proportional to $r^{-4}$ is present,\footnote{This is not highly surprising as, in the vacuum case, the $r^{-4}$ term is proportional to $\kappa_0$, that arise as an integration constant of the equation for $B$~\eqref{eq:static_eom_B}. So if one imposes $B=0$ at the beginning, there will be no  $r^{-4}$ term in vacuum also.} so that matter does not source any $r^{-4}$ term in the function $F(r)$. Then for this interior solution for which $B=0$, we can set at its border, i.e.\ at the radius of the star, $r=r_*$  (where $\rho(r_*)=0$), the matching conditions with the exterior 3D vacuum metric. On doing this we find that also in the exterior metric, we need to set $B=0$, and in particular $\kappa_0=0$. On the other hand, the exterior solution fixes $\Lambda_0=0$ also for the interior solution. Since $B=0$ also for the exterior solution, this solution represents a static asymptotically flat solution. In other words, we reach the conclusion, that, on choosing $B=0$ for the interior matter static solution, the 4D metric solution surrounding a spherically symmetric star, reduces to the GR standard Schwarzschild metric.

For the general case $B\neq0$, on the other hand, since we do not know the interior solution in general, in turn, we are not able to use the matching conditions between the interior and exterior solutions to determine the value of $\kappa_0$ analytically. Instead, one should set the value of $\kappa_0$ by looking for the numerical interior solution for the star for a given equation of state, after setting appropriate boundary conditions at the origin, and then matching the interior solution to the exterior solution.

\section{Time-dependent solutions}\label{sec:time_dep}

Let us consider now vacuum time-dependent solutions for the equations of motion. The gauge-fixing constraint imposes that $\phi(r,t) = \phi(t)$. One can then solve $\mathcal{E}_{N}=0$ (i.e.\ the equation of motion for the lapse) for $B$, and
then $\mathcal{E}_{B}=0$ (i.e.\ the equation of motion for the shift) for $F$ to find 
\begin{equation}
F=\left(1-\frac{2\mu(t)}{r}-\frac{1}{3}\,\Lambda(t)\,r^{2}+\frac{\kappa^{2}(t)}{r^{4}}\right)^{-1/2},
\end{equation}
where we have fixed $F>0$. Here $\mu(t)$ and $\kappa(t)$ are two constants (in $r$) of integration and we have defined $\Lambda(t) = V[\phi(t)] - \phi^2(t)/3$.
Note the similarity with the static case~\eqref{eq:N_sol}.
Then on fixing $F$ to such a solution, one finds that, on reconsidering $\mathcal{E}_{N}=0$, one can solve it for $N$, finding the following equation
\begin{equation} \label{eqn:Nbranches}
N=\frac{r^{2}\,(rB_{,r}-B-rF_{,t})}{3\kappa F}\,.
\end{equation}

This solution for $N$ is valid unless $\kappa=0$. The case for which $\kappa=0$ will be treated separately in the appendix. Appropriate boundary conditions need to be set for the solution so that, for instance, $N$ remains finite in the wanted coordinate patch (in particular at infinity). These same boundary conditions will take care of the case $0 < |\kappa| \ll 1$. 
At this point, both the equations $\mathcal{E}_{N}=0$, and $\mathcal{E}_{B}=0$ are satisfied, and the equation of motion $\mathcal{E}_{rr}=0$, can be used in order to solve for the $B$ field.  We find

\begin{eqnarray} \label{eqn:Bsecondbranch}
B & = & \frac{b_{1}(t)}{r^{2}}+r\,b_{2}(t)+\frac{r}{6}\,\int^{r}\frac{\Omega(t,r')}{(r')^{5}}\,dr'-\frac{1}{6r^{2}}\,\int^{r}\frac{\Omega(t,r')}{(r')^{2}}\,dr'\,,\\ \label{eq:B_time_dep}
\Omega & = & -2\kappa\dot{\phi}F^{3}r^{3}+F^{3}\,(r^{6}\Lambda_{,\phi}\dot{\phi}+12\kappa\dot{\kappa})+3F^{2}F_{,t}\,(r^{6}\Lambda-r^{4}+3\kappa^{2})+7r^{4}F_{,t}\,,\label{eq:Omega_def}
\end{eqnarray}
where the functions $b_{1,2}$ are free functions which come as integration constants (i.e.\ constant in $r$) for the two integrals defining the solution for $B$.

We will now proceed to understand the properties of such a solution. First of all we get the following relations
\begin{eqnarray}
R & = & \frac{2}{r^{2}}-\frac{2}{r^{2}F^{2}}+\frac{4F_{,r}}{rF^{3}}=2\Lambda(t)+\frac{6\kappa^{2}}{r^{6}}\,,\label{eq:time_var_R}\\
K & = & -\frac{3}{r}\left(\frac{B}{NF}+\frac{\kappa}{r^2}\right)\,,\\
K^{ij}K_{ij} & = & \frac{3}{r^2}\left(\frac{B}{NF}+\frac{\kappa}{r^2}\right)^2+\frac{6\kappa^2}{r^6}\,,\label{eq:Kij_sq}\\
\lambda & = &\frac{2}{r}\left(\frac{B}{NF}+\frac{\kappa}{r^2}\right)-\frac{2}{3}\,\phi\,,\\
\lambda^r & = & \frac{\ell(t)}{r^{2}}+\frac{1}{r^{2}}\int^{r}NF\varrho^2\left[\Lambda_{,\phi}+\frac{2}{\varrho}\left(\frac{B}{NF}+\frac{\kappa}{\varrho^2}\right)\right]d\varrho\,.
\end{eqnarray}
Here $\ell(t)$ is a constant (in $r$) of integration.
When $\kappa\neq0$ the 3D Ricci scalar only blows up at the origin $r=0$. In this case we see that a singularity is present at the origin in general. Therefore we need a horizon to make sure it is not a naked one.

\subsection{Coordinate singularity}

We will now first investigate the apparent singularity at $r=r_{0}(t)$, defined by the property 
\begin{equation}
\frac{1}{F(r_{0})^{2}}=0\,,
\end{equation}
which can be formally solved for $\mu(t)$ as 
\begin{equation}
\mu(t)=\frac{r_{0}}{2}+\frac{\kappa^{2}}{2r_{0}^{3}}-\frac{1}{6}\,\Lambda r_{0}^{3}\,.\label{eq:mu_sz_sing}
\end{equation}
Then if we write $r=r_{0}(t)+\rho$, and assume $0<\rho\ll r_{0}$,
then we find 
\begin{equation}
\frac{1}{F^{2}}=\frac{r_{0}^{4}-3\kappa^{2}-r_{0}^{6}\Lambda}{r_{0}^{5}}\,\rho+\mathcal{O}(\rho^{2})\,,
\end{equation}
which leads to the condition $r_{0}^{4}-3\kappa^{2}-r_{0}^{6}\Lambda\geq0$.
It is easy to show, that in this case the proper distance, $\int F\text{d}r$, remains finite around this point. On calculating the limit of several expressions around this critical point, $r=r_{0}(t)$, we find the following results
\begin{eqnarray}
\lim_{r\to r_{0}^{+}}\lambda & = & -\frac{3(\Gamma_{4}-2\kappa\tilde{\Gamma}_{5})+2\tilde{\Gamma}_{5}\phi r_{0}^{3}}{3r_{0}^{3}\tilde{\Gamma}_{5}}\,,\\
\lim_{r\to r_{0}^{+}}\lambda^r & = & \frac{\ell(t)}{r_{0}^{2}}\,,\\
\lim_{r\to r_{0}^{+}}K & = & \frac{3\,(\Gamma_{4}-2\kappa\tilde{\Gamma}_{5})}{2\tilde{\Gamma}_{5}r_{0}^{3}}\,,\\
  \lim_{r\to r_{0}^{+}}K_{ij}K^{ij} & = & \frac{3(\Gamma_{4}^{2}-4\kappa\tilde{\Gamma}_{5}\Gamma_{4}+12\kappa^{2}\tilde{\Gamma}_{5}^{2})}{4\tilde{\Gamma}_{5}^{2}r_{0}^{6}}\,,\\
  \Gamma_{4} & = & \Gamma_{4}(t)\equiv r_{0}^{6}\Lambda_{,\phi}\dot{\phi}+6\dot{\mu}\,r_{0}^{3}-6\kappa\dot{\kappa}\,,\\
\tilde{\Gamma}_{5} & = & \tilde{\Gamma}_{5}(t)\equiv r_{0}^{3}\dot{\phi}-3\dot{\kappa}\,.
\end{eqnarray}
All the scalars remain finite at the critical points $r=r_{0}(t)$ (we recall that the Ricci scalar~\eqref{eq:time_var_R} is finite everywhere except at the origin $r=0$), so that we can conclude that this point corresponds only to a coordinate singularity. We will focus since now on, on the presence of horizons for this metric.

\subsection{Trapping horizons and null surfaces}

In order to study the presence of horizons for an effective 4D metric defined as in Eq.~\eqref{eq:ansatz}
\begin{equation}
\text{d}s^{2}=-(N^{2}-B^{2})\,\text{d}t^{2}+2BF\,\text{d}t\text{d}r+F^{2}\text{d}r^{2}+r^{2}\text{d}^{2}\Omega\,,
\end{equation}
it is convenient to introduce two future-oriented four-vectors 
\begin{equation}
l^{\mu} = \frac{1}{F}\,(F,N-B,0,0)\,,\qquad
n^{\mu} = \frac{1}{F}\,(F,-N-B,0,0)\,,\label{eq:null_vect}
\end{equation}
$l^\mu$ being ``outgoing'' and $n^\mu$, ``ingoing.''
Those vectors are null and conveniently normalized, namely
\begin{equation}
n^{\mu}n_{\mu} = 0\,, \qquad
l^{\mu}l_{\mu} = 0\,,\quad
n^{\mu}l_{\mu} = -2N^{2}\,.
\end{equation}
Out of these two light-like vectors, on following e.g.\ \cite{Faraoni:2013aba}, we can then build up a projector 
\begin{equation}
h^{\mu\nu}=g^{\mu\nu}+\frac{l^{\mu}n^{\nu}+l^{\nu}n^{\mu}}{(-n^{\alpha}l_{\alpha})}\,,
\end{equation}
and we can introduce the corresponding expansion scalars
\begin{equation}
\theta_{l} = h^{\mu\nu}\,\nabla_{\mu}l_{\nu}=\frac{2}{rF}\left(N-B\right),
\qquad
\theta_{n} = h^{\mu\nu}\,\nabla_{\mu}n_{\nu}=-\frac{2}{rF}\left(N+B\right).
\end{equation}
In terms of the expansions, we have an expression of $g^{rr}$ as
\begin{equation}
g^{rr}=\frac{1}{F^{2}}-\frac{(\gamma^{rr}N_{r})^{2}}{N^{2}}=\frac{N^{2}-B^{2}}{N^{2}F^{2}} = -\frac{1}{4N^2}\theta_{l}\theta_{n}\,. \label{eqn:grr-timedependent}
\end{equation}
We now define $r_H(t)$ by 
\begin{equation}
B\bigl(t,r=r_{H}(t)\bigr)=N\bigl(t,r=r_{H}(t)\bigr)\,,\label{eq:MOTS}
\end{equation}
and $S_{H}$ by 
\begin{equation}
S_{H}\equiv r-r_{H}(t)=0\,.
\end{equation}
Then the surface $S_{H}=0$ with $B>0$ and $0<F<\infty$ represents a marginally outgoing trapped surface (MOTS), as $\theta_{l}=0$ and $\theta_{n}<0$ there. This surface represents an apparent horizon as (\ref{eqn:grr-timedependent}) implies $g^{rr}=0$ on the MOTS.

Instead the surface representing the coordinate singularity $S_{0}\equiv r-r_{0}(t)=0$, with $r_{0}$ satisfying $F\bigl(r=r_{0}(t)\bigr)\to\infty$, is then trapped (since both $\theta_l$ and $\theta_n$ are negative). In fact, for this $S_0$ surface, we find that both $N$ and $B/F$ get the following values
\begin{eqnarray}
\lim_{r\to r_{0}^{+}}N & = & \frac{2}{3}\frac{r_{0}^{3}\,\tilde{\Gamma}_{5}}{r_{0}^{4}-r_{0}^{6}\Lambda-3\kappa^{2}}\,,\label{eq:N_0_1st}\\
\lim_{r\to r_{0}^{+}}\frac{B}{F} & = & -\frac{r_{0}}{3}\,\frac{\Gamma_{4}}{r_{0}^{4}-r_{0}^{6}\Lambda-3\kappa^{2}}\,,\label{eq:B_0_2nd}
\end{eqnarray}
which sets $\tilde{\Gamma}_{5}\geq0$.

The two surfaces $S_0=0$ and $S_H=0$ are identical only if $B=N$ and $1/F \rightarrow 0$. The conditions~\eqref{eq:N_0_1st}-\eqref{eq:B_0_2nd} then impose that $\Gamma_4 = 0$, which in turn imposes that $r_0$ (and thus $r_H$) is constant in time, as can be seen by deriving Eq.~\eqref{eq:mu_sz_sing}. In this case the expansion scalars are vanishing at the surface, thus the MOTS condition degrades to a marginally trapped surface (MTS) one. Note that this property is strongly linked to the static case, where $N = F^{-1}$, thus $\theta_l$ and $\theta_n$ vanish on $S_0$, which is then a MTS.

Otherwise, when the two surfaces are distinct, the continuity of $\theta_l$ implies that $r_0(t) < r_H(t)$, i.e.\ that the coordinate singularity lies inside the MOTS.

\subsection{Type of the horizons}

Let us consider the MOTS, $S_{H}=0$, and study whether it is spacelike or not. Since $g^{rr}$ vanishes on this surface, we find 
\begin{equation}
g^{\mu\nu}\partial_{\mu}S_{H}\partial_{\nu}S_{H}=-\frac{\dot{r}_{H}^{2}}{N^{2}}-\frac{2\dot{r}_{H}}{NF}\,,
\end{equation}
so that this surface is spacelike if $0<-\dot{r}_{H}<2N/F\bigl(t,r_{H}(t)\bigr)$. Taking the time derivative of Eq.~\eqref{eq:MOTS}, we can express $\dot{r}_{H}$, so that $S_H$ is spacelike if
\begin{equation}
0 < \left. \frac{B_{,t}-N_{,t}}{B_{,r}-N_{,r}}\right\vert_{r=r_H} < \left. \frac{2 N}{F}\right\vert_{r=r_H}\,,
\end{equation}
and lightlike if one of the inequalities is saturated.

As for the coordinate singularity surface $S_{0}=r-r_{0}(t)=0$ (with $r_{0}(t)>0$ and $F\bigl(r=r_{0}(t)\bigr)\to\infty$), we have
\begin{equation}
g^{\mu\nu}\partial_{\mu}S_{0}\partial_{\nu}S_{0}  =
 -\frac{\dot{r}_{0}^{2}}{N^{2}}-\frac{2\gamma^{rr}N_{r}}{N^{2}}\,\dot{r}_{0}+\gamma^{rr}-\frac{(\gamma^{rr})^{2}(N_{r})^{2}}{N^{2}}
 =-\frac{1}{N^{2}}\left(\dot{r}_{0}+\frac{B}{F}\right)^{2}+\frac{1}{F^{2}}=0\,.
\end{equation}
To obtain the last equality, we need to take the time derivative of Eq~\eqref{eq:mu_sz_sing}, and use Eq.~\eqref{eq:B_0_2nd}, which leads to $\dot{r}_{0}=-\lim_{r\to r_{0}}B/F$, so that $S_{0}=0$ is always a null surface.

\subsection{A possible violation of the null convergence condition}

Let's recall that in GR the null energy condition for the stress energy tensor of matter and the null convergence condition for the geometry are equivalent. The latter states that for every future directed null vector $l^\mu$, $R_{\mu\nu}l^{\mu}l^{\nu} \geq 0$. In non-GR gravitational theories the null energy condition does not necessarily implies the null convergence condition. In our theory, we shall see below that the null convergence condition can in principle be violated even in the absence of matter. For simplicity we study the possible violation of the null convergence condition on the MOTS ($S_{H}=0$) and on the $S_0=0$ surface.

In our theory, using the ``outgoing'' null vector~\eqref{eq:null_vect}, we have 
\begin{equation}
R_{\mu\nu}l^{\mu}l^{\nu} =
\frac{2\left(N-B\right)^2}{F^2r}\left(\frac{F_{,r}}{F}+\frac{N_{,r}}{N}\right)+\frac{4\left(N-B\right)F_{,t}}{F^2r}+ \frac{2B}{Fr}\left(\frac{B_{,t}}{B}-\frac{N_{,t}}{N}\right)\,.\label{eq:Rll_gen}
\end{equation}
On the MOTS, on imposing the condition $B=N$, we simply find
\begin{equation}
R_{\mu\nu}l^{\mu}l^{\nu}=\frac{2}{r_{H}F}\,(B_{,t}-N_{,t})|_{r=r_{H}}=\frac{2}{F}\,\frac{\dot{r}_{H}}{r_{H}}\,(N_{,r}-B_{,r})|_{r=r_{H}}\,.\label{eq:Rll}
\end{equation}
In order to proceed, we need to study better the surface $S_{H}=0$. It is possible to solve the relation
\begin{equation}
B(t,r_{H}(t))=N(t,r_{H}(t))\,,
\end{equation}
for example in terms of $b_{2}(t)$. On taking this solution and inserting it into Eq.\ (\ref{eq:Rll}), we find a complicated relation which does not need to be non-negative, in principle. Thus the null convergence condition can be violated in our theory in the absence of matter. 

One can arrive at a similar conclusion on the coordinate singularity surface $S_{0}=r-r_{0}(t)=0$. We can consider a light-like vector $l^{\mu}$ evaluated on the $S_0=0$ surface. In this case one can choose $l^{\mu}$ so that $\lim_{r\to r_{0}}l^{\mu}$ still gives a non-trivial vector whose components are not blowing up at the surface. In this case, one can evaluate Eq.~\eqref{eq:Rll_gen} on the $S_{0}$ surface. On doing so, one finds that the limit is finite (as no physical singularity is expected to be at $r=r_0(t)$), but this limit gives a rather complicated expression which depends, in particular on the second time derivative of the free functions $\mu$, $\kappa$, $\phi$, and does not necessarily need to be non-negative.

\subsection{Matching conditions with cosmology}

From what we have said in the case of static solution, we match the vacuum time-dependent solutions with a vacuum FLRW cosmology solution at infinity. In other words, we impose
\begin{equation}
\lim_{r\to\infty} R = 0\,,\quad \lim_{r\to\infty} K = 3H_0\,, \quad \lim_{r\to\infty} K^{ij}K_{ij} = 3H_0^2\,, \quad \lim_{r\to\infty} \lambda^r = 0\,, \quad \dot{\phi} = 0\,,  \label{eqn:dS-boundarycondition}
\end{equation}
where $H_0$ is a positive constant. The last condition comes from the fact that, in vacuum FLRW cosmology, $\rho_\phi$ has to be constant, see~\eqref{eqn:Friedman}. We also demand that the mass of the compact object be finite. A well-known definition of the mass in an asymptotically de Sitter spacetime is the Abbott-Deser mass~\cite{Abbott:1981ff}, which is know to be conserved. In spherical symmetry, the Abbott-Deser mass reduces to the $r\to\infty$ limit of the generalized Misner-Sharp mass $M$ defined in (\ref{eqn:def-Misner-Sharp-mass}) with $\Lambda_{\text{eff}}=3H_0^2$. Therefore, we require that 
\begin{equation}
 h^{ab}\partial_a r\partial_b r = 1 - H_0^2r^2 + \mathcal{O}(r^{-1})\,. \label{eqn:finite-ADmass}
\end{equation}

On demanding the first and the last condition in (\ref{eqn:dS-boundarycondition}), we find that we need to set 
\begin{equation}
 \Lambda=0\,, \quad \dot\phi=0\,,
\end{equation}
so that we can set
\begin{equation}
F=\left(1-\frac{2\mu(t)}{r}+\frac{\kappa^{2}(t)}{r^{4}}\right)^{-1/2}\,.
\end{equation}
We then demand the second condition in (\ref{eqn:dS-boundarycondition}) to find
\begin{equation}
 b_2 = \frac{H_0}{2\kappa}(5\mu^2\dot{\mu}+2b_1)\,,
\end{equation}
which ensures the third condition as well. Here, we assume $\kappa\ne 0$. The case with $\kappa=0$ will be considered separately in appendix. So far the free function $b_1$ is left unspecified. In order to simplify the expressions, we can redefine it as
\begin{equation}
  b_1\equiv -n(t)\kappa-\frac52\,\dot\mu\,\mu^2\,,\qquad\textrm{so that}\qquad
  \lim_{r\to\infty}N=n(t)>0\,,
\end{equation}
to match a yet undetermined and free behavior at infinity\footnote{We can e.g.\ choose $n=1$ by means of the freedom of the space-independent time reparametrization, which is a part of the foliation-preserving diffeomorphism (\ref{eqn:foliation-preserving-diffeo}).}. In this case $b_2=-H_0\,n(t)$.

The forth condition then leads to
\begin{equation}
  n\,(2H_0-\Lambda_{,\phi})=0\,,
\end{equation}
which reduces to the same condition we have found in the static case, namely $2H_0-V_{,\phi}+2\phi/3=0$.
 
Finally, demanding the finite mass condition (\ref{eqn:finite-ADmass}), one obtains
\begin{equation}
 \dot{\mu} + H_0\dot{\kappa} = 0\,.  \label{eqn:finitemasscondition}
\end{equation}

After imposing all the above mentioned boundary conditions the $4$-dimensional Ricci tensor is shown to be
\begin{equation}
R_{\mu\nu}=\Lambda_\text{eff}\, g_{\mu\nu}\,, \quad  \Lambda_\text{eff} = 3H_0^2 \,, \label{eqn:Rmunu-timedependentcase}
\end{equation}
where $\Lambda_{\text{eff}}$ is an effective cosmological constant. Therefore, from a GR point of view, the effective $4$-dimensional metric is locally Schwarzschild ($H_0=0$) or Schwarzschild-AdS ($H_0\ne 0$), again thanks to the Birkhoff theorem. By using the definition (\ref{eqn:def-Misner-Sharp-mass}), we find that the generalized Misner-Sharp mass is 
\begin{equation}
 M = \mu + H_0\kappa\,, \label{eqn:M-timedependentcase}
\end{equation}
which is constant due to the condition (\ref{eqn:finitemasscondition}).  Note again the similarity with the static case (\ref{eq:expr-Misner-Sharp-mass}). This is consistent with the fact that the $4$-dimensional metric is either the Schwarzschild ($H_0=0$) or Schwarzschild-AdS ($H_0\ne 0$) metric. Therefore, the Abbott-Deser mass $\lim_{r\to\infty}M$ is constant and finite. 

Similar results are found for the particular case for which $\kappa=0$, as shown in appendix.

\subsection{Exterior solution of a time-dependent star}

While the combination (\ref{eqn:M-timedependentcase}) is constant, each of $\mu(t)$ and $\kappa(t)$ may depend on time and one of them can be thought of being a free function of the solution. Nonetheless the theory is free from extra local degrees of freedom and indeed the functions $\mu(t)$ and $\kappa(t)$ are completely determined by matching the interior with the exterior vacuum solution. The whole system including the functions $\mu(t)$ and $\kappa(t)$ in the vacuum region are unambiguously determined by the boundary condition and the initial condition of dynamical degrees of freedom. 

For the interior solution of a star, we need to impose the boundary conditions which are compatible with the absence of a singularity at the origin. Through the matching conditions at the border of the compact object, these boundary conditions will in turn affect the exterior solution. On the other hand, the boundary conditions imposed by cosmology will also affect the interior solution by fixing the free functions which cannot be set by the boundary conditions at the origin. For example, around the origin, we will have in general $B=\zeta_1(t)\,r+\mathcal{O}(r^2)$, and $\zeta_1(t)$ will be determined by the interplay between the dynamics of matter and the matching conditions with the exterior solution.

\section{Summary and discussion}\label{sec:summary}

We have studied spherically symmetric solutions of a type-II MMG theory considering both cases of time-independent and time-dependent metrics. The theory itself was constructed in a previous paper~\cite{DeFelice:2020eju} and was proven to have only two local physical degrees of freedom but does not possess an Einstein frame. The main goal of the present paper was to further explore implications and predictions of the theory for strong gravitational systems, i.e.\ black holes.

In the static case, a solution is characterized by five parameters as in Eq. (\ref{eq:static_sol}), and represents either a Schwarzschild or Schwarzschild-(A)dS metric. What is interesting here is that the effective cosmological constant deduced from the curvature of the static metric depends not only on the gravity action but also on the way the spacetime is foliated by constant-time hypersurfaces (this is a natural consequence of the breaking of temporal diffeomorphism). The resulting physical quantities from this metric, including the intrinsic and extrinsic curvatures of the constant time hypersurfaces, show that there are no physical singularities except at the origin of the spherical symmetry. 

In the time-dependent case, despite the absence of extra local physical degrees of freedom, the theory admits non-trivial solutions that are not Einstein spaces in general and that possess regular trapping and event horizons. The solutions are parametrized by several functions of time and are in general different from Schwarzschild or Schwarzschild-(A)dS solutions even locally. The time-dependent solution can in principle break the null convergence condition despite the fact that the solution is in vacuum and thus without exotic matter. 

The situation becomes simpler if we require that the solution should asymptotically approach the corresponding cosmological solution, i.e.\ the de Sitter (or Minkowski) spacetime in the spatially homogeneous and isotropic slicing. In a more realistic situation, the boundary condition within the spherically symmetric ansatz may/should be modified one way or another to the extent that the would-be mismatch between the cosmological solution and the black hole solution can be absorbed by physical effects in the intermediate region (such as deviations from spherical symmetry, the existence of interstellar matter, etc.) and/or by non-trivial cosmology (e.g. the existence of cold dark matter, baryons, radiation, etc.). Nonetheless, the properties of the black hole solution at astrophysical scales are expected to be insensitive to such modifications of the boundary condition as far as there is a large enough hierarchy between the size of the black hole and the scales associated with the physical effects in the intermediate and cosmological scales. For this reason we have considered the idealized situation where the spherically symmetric solution asymptotically matches with the de Sitter (or Minkowski) spacetime in the spatially homogeneous and isotropic slicing. 

For the static case, the cosmological boundary condition completely eliminates the mismatch between the effective cosmological constant deduced from the solution and that from the action: the effective cosmological constant agrees with the cosmological value that is determined by the gravitational action. Besides the effective cosmological constant $\Lambda_\text{eff}$ and the black hole mass $M$, there is one additional parameter $\kappa_0$, representing the way the spacetime is foliated by constant time hypersurfaces. Nonetheless, since $\kappa_0$ is independent from $\Lambda_\text{eff}$ and $M$, it does not affect the covariant properties of the $4$-dimensional metric. Therefore, any astrophysical processes of matter cannot probe the value of $\kappa_0$ if the matter action is invariant under the spacetime diffeomorphism. The only way to probe $\kappa_0$ is through gravity, e.g. gravitational waves. The value of $\kappa_0$ should be determined by a boundary condition. For example, if the static vacuum solution is to represent an external solution of a static star, then $\kappa_0$ should be determined by a matching condition with an interior static solution with a regular center. On the other hand, if the solution is to represent a static black hole formed by gravitational collapse then $\kappa_0$ should be obtained by the dynamics of gravitational collapse starting from a regular initial condition. 

For the time-dependent case, even after imposing the cosmological boundary condition, the solutions in general still deviate from those in GR, i.e.\ they are not Einstein spaces in general. However, for those non-GR solutions the Abbott-Deser mass diverges and thus they do not represent compact objects. Such solutions may be interesting as possible large-scale structures in the universe but are not suitable for astrophysical compact objects. 

Since our main focus in the present paper was on compact objects at astrophysical scales, we required that the Abbott-Deser mass be finite. This condition eliminates the non-GR time-dependent solutions and thus only possible solutions in the setup turned out to be Schwarzschild or Schwarzschild-(A)dS ones even in the time-dependent case. 

Despite the absence of extra physical degrees of freedom, we have seen that the Birkhoff theorem does not hold in the type-II MMG theory and that a spherically symmetric vacuum solution depends not only on the cosmological constant and the mass but also on several free functions of the globally defined time coordinate. Those free functions are specified by suitable asymptotic conditions at the spatial infinity or boundary conditions at a finite distance. This suggests the existence of modes satisfying elliptic (instead of hyperbolic) equations. Such modes are often called instantaneous modes or shadowy modes, and their behavior is controlled by boundary conditions (instead of initial conditions)~\cite{DeFelice:2018ewo}. It is expected that those modes should be useful in understanding apparent discrepancy with some results in the literature~\cite{Khoury:2013oqa,Pajer:2020wnj}. It is therefore worthwhile investigating properties of the instantaneous/shadowy modes in the context of the type-II MMG theory. 

In the present paper we have studied vacuum solutions in the type-II MMG theory. In order to understand the structure of compact objects in the strong gravity regime more deeply, we need to extend the analysis to the system with matter. It is also intriguing to study perturbations around the spherically symmetric solutions to understand the stability of the system and the properties of gravitational waves. Also, it is certainly interesting to formulate the initial value problem in general and then to perform numerical simulations of gravitational collapse.

\begin{acknowledgments}
The work of A.D.F.\ was supported by Japan Society for the Promotion of Science Grants-in-Aid for Scientific Research No.~20K03969. The work of S.M.\ was supported in part by Japan Society for the Promotion of Science Grants-in-Aid for Scientific Research No.~17H02890, No.~17H06359, and by World Premier International Research Center Initiative, MEXT, Japan. 
\end{acknowledgments}

\appendix*

\section{Case $\kappa=0$}\label{sec:app}

Let us consider the solution for $F(t,r)$ in the case $\kappa=0$, so that
\begin{equation}
  F=\left( 1-\frac{2\mu(t)}r - \frac13\, \Lambda(t)\,r^2 \right)^{\!-1/2},
\end{equation}
then in this case, we can show that the equation $\mathcal{E}_N=0$ reduces to
\begin{equation}
  \frac{\Mpl^2}{3r^2N^2F^2}\,(rF_{,t}-rB_{,r}+B)^2=0\,,
\end{equation}
which is solved by
\begin{equation}
  B=b_2(t)\,r+r\int^r \frac{1}{r'}\,\frac{\partial F(t,r')}{\partial t}\,dr'\,.
\end{equation}
At this moment both equations $\mathcal{E}_N=0$ and $\mathcal{E}_B=0$ are satisfied. The remaining independent equation of motion, namely $\mathcal{E}_{rr}=0$, can be used to solve for the variable $N$ as follows
\begin{equation}
  N=\frac{n(t)}{F}-\frac{{\dot\phi}}{3F}\int^r r' F(t,r')^3\, dr'\,.
\end{equation}
On the other hand, the boundary conditions at infinity impose that $\Lambda(t)=0=\dot\phi$. For such a case we find that the condition $\lim_{r\to\infty}K=3H_0$ leads to
\begin{equation}
  b_2=-H_0\,n(t)+\frac{\dot\mu}\mu\,.
\end{equation}
On considering solutions describing compact objects, one imposes that $h^{ab}\partial_a r\partial_b r=1-H_0^2r^2+\mathcal{O}(r^{-1})$, which in turn gives the condition
\begin{equation}
  \dot\mu=0\,.
\end{equation}

In this case $F$ becomes time independent. Finally, on setting $n(t)=1$, by means of the freedom of the space-independent time reparametrization, which is a part of the foliation-preserving diffeomorphism (\ref{eqn:foliation-preserving-diffeo}), we reduce then to the static case for $\kappa_0=0$, that is
\begin{eqnarray}
  F&=&\left( 1-\frac{2\mu_0}r \right)^{\!-1/2}\,,\\
  B&=&-H_0\, r\,,\\
  N&=&\frac1{F}\,.
\end{eqnarray}

\end{document}